\documentclass[twocolumn,noshowpacs,preprintnumbers,superscriptaddress,amsmath,amssymb,floatfix, groupedaddress,showkeys,showpacs]{revtex4}
\usepackage[caption=false]{subfig}
\usepackage{graphicx}
\usepackage{color}
\usepackage{setspace}
\usepackage{placeins}
\usepackage{tikz}
\usepackage{mathtools}
\usepackage{amsthm}
\usepackage{color}
\usepackage{xkeyval,xcolor}
\usepackage{array}

\begin{document}

\title{Spatio-temporal propagation of COVID-19 pandemics}
\date{\today}
\author{Bnaya Gross} 
\email{bnaya.gross@gmail.com}
\affiliation{Department of Physics, Bar-Ilan University, Ramat-Gan 52900, Israel}
\author{Zhiguo Zheng} 
\affiliation{School of Reliability and Systems Engineering, Beihang University, Beijing 100191, China}
\author{Shiyan Liu} 
\affiliation{School of Reliability and Systems Engineering, Beihang University, Beijing 100191, China}
\author{Xiaoqi Chen} 
\affiliation{School of Reliability and Systems Engineering, Beihang University, Beijing 100191, China}
\author{Alon Sela} 
\affiliation{Department of Industrial Engineering, Ariel University, Ariel, Israel}
\affiliation{Department of Physics, Bar-Ilan University, Ramat-Gan 52900, Israel}
\author{Jianxin Li} 
\affiliation{Beijing Advanced Innovation Center for Big Data and Brain Computing, Beihang University, Beijing 100083, China}
\affiliation{State Key Laboratory of Software Development Environment, Beihang University, Beijing 100083, China}
\author{Daqing Li} 
\affiliation{Beijing Advanced Innovation Center for Big Data-Based Precision Medicine, Beihang University, Beijing 100191, China}
\affiliation{School of Reliability and Systems Engineering, Beihang University, Beijing 100191, China}
\author{Shlomo Havlin} 
\affiliation{Department of Physics, Bar-Ilan University, Ramat-Gan 52900, Israel}
\date{\today}

\begin{abstract}
	The new coronavirus known as COVID-19 is spread world-wide since December 2019. Without any vaccination or medicine, the means of controlling it are limited to quarantine and social distancing. Here we study the spatio-temporal propagation of the first wave of the COVID-19 virus in China and compare it to other global locations. We provide a comprehensive picture of the spatial propagation from Hubei to other provinces in China in terms of distance, population size, and human mobility and their scaling relations. Since strict quarantine has been usually applied between cities, more insight about the temporal evolution of the disease can be obtained by analyzing the epidemic within cities, especially the time evolution of the infection, death, and recovery rates which affected by policies. We study and compare the infection rate in different cities in China and provinces in Italy and find that the disease spread is characterized by a two-stages process. At early times, at order of few days, the infection rate is close to a constant probably due to the lack of means to detect infected individuals before infection symptoms are observed. Then at later times it decays approximately exponentially due to quarantines. The time evolution of the death and recovery rates also distinguish between these two stages and reflect the health system situation which could be overloaded.
\end{abstract}

\maketitle

\section{Introduction}
Since December 2019 the world is fiercely struggling against an epidemics of a novel Coronavirus named COVID-19 identified in Wuhan, a city of 11 million people in Hubei Provence, China. A medical cure from the disease is yet unavailable and the number of infected cases is increasing. As for April 1 2020, the virus has already spread to more than 100 countries around the world with more than 1 million confirmed cases. \par
\begin{figure}
	{\includegraphics[scale=0.55]{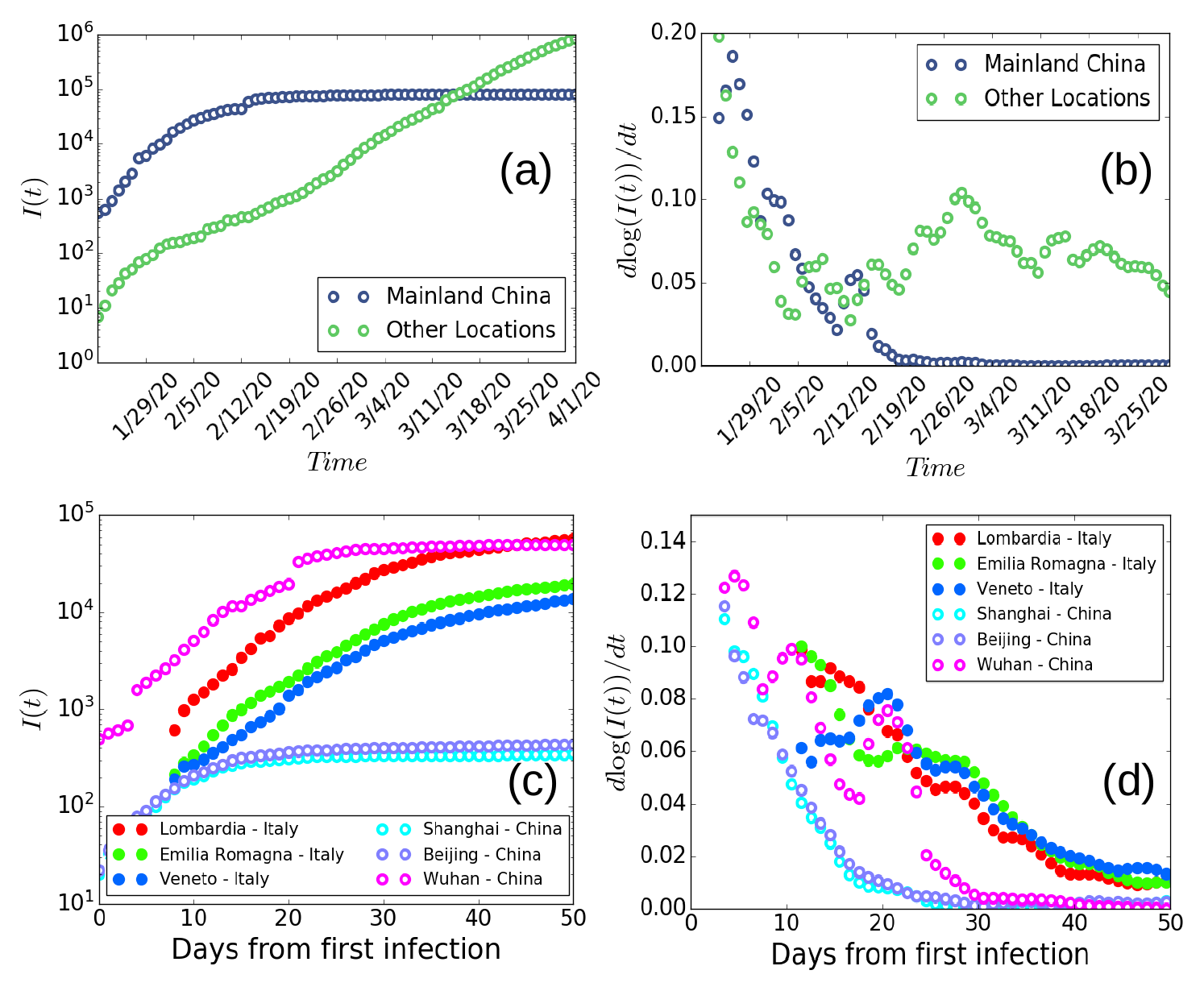}}
	\caption{\textbf{General view of the COVID-19 propagation as for April 1, 2020.} \textbf{(a)} The number of confirmed infected cases $I(t)$ in mainland China and other locations around the globe on a semi-log scale. Less than two months since COVID-19 inception, the number of infected individuals in China almost reaches saturation while the number in other locations have been rapidly increasing. \textbf{(b)} The slope (derivative) of $\log(I(t))$. The COVID-19 propagation decays rapidly in most cities in China and the disease is almost stabilized with the derivative approaches zero. In contrast, cities in other locations around the globe have been still in their early stage and the disease was still spreading as can be seen by the almost constant or even increase of the derivative. \textbf{(c)}  The number of confirmed infected cases $I(t)$ in different locations in China and Italy since the first infection. While in most cities in China the disease stabilized after a short time of approximately 20 days on average, the disease in Italy has been still spreading and approached stability much later, see  \textbf{(d)}. The slope (derivative) of $\log(I(t))$ in different locations in China and Italy since the first infection.}
	\label{fig:global_cases}
\end{figure}

%\textbf{(c)} Simulations results of Gaussian random walker (left) compared to L\'{e}vy flight random walker (right). The disease spread using long-hops of the human mobility  L\'{e}vy flight, when no quarantine measures are applied.
\begin{figure*}
	\begin{center}
	\includegraphics[scale=0.516]{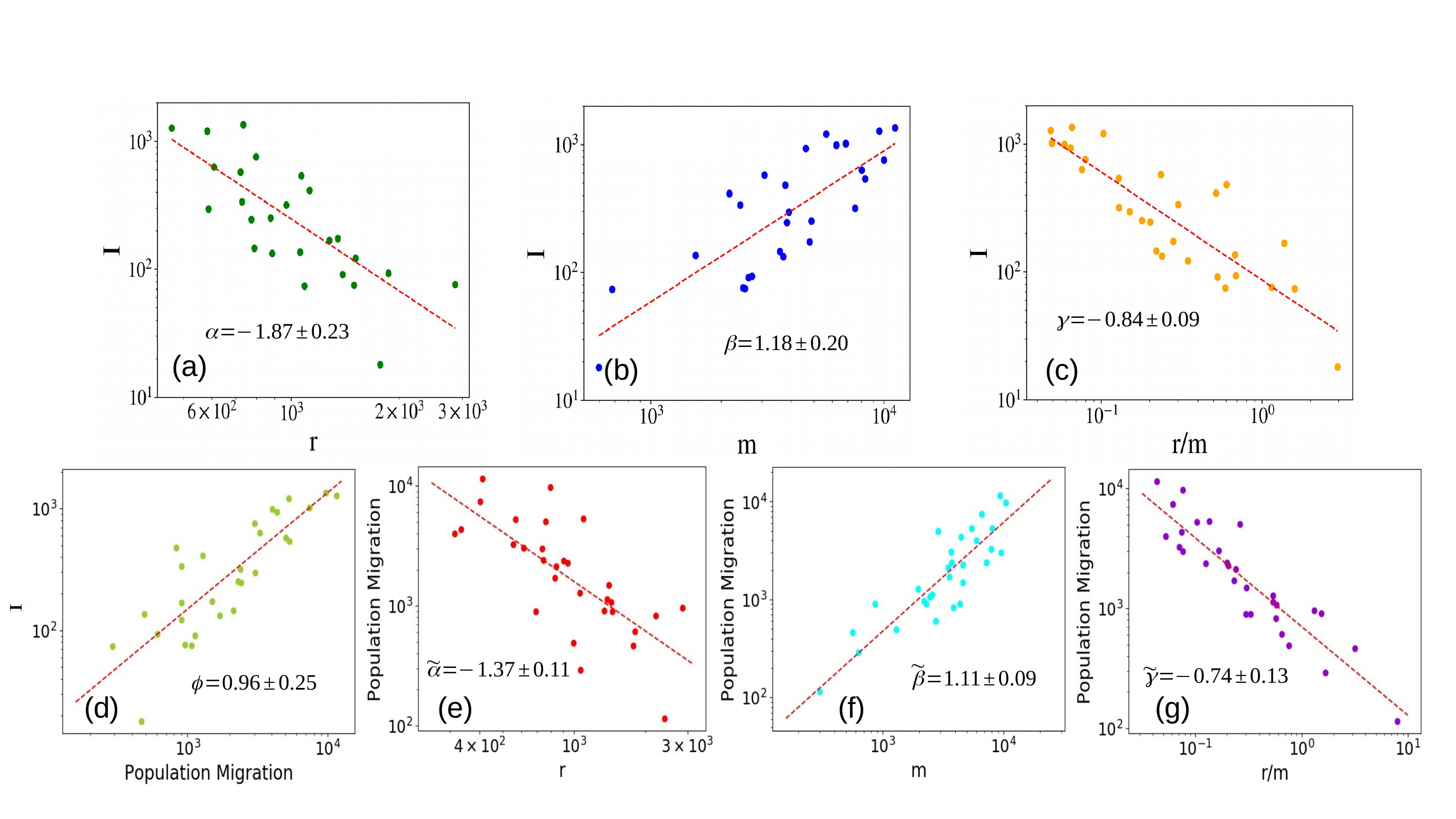}	
	\caption{\textbf{Spatial propagation analysis of the COVID-19 in China.} \textbf{(a)} The number of infected individuals $I$ (as of  March 1, 2020) as a function of the distance from Hubei. The scaling follows Eq.  \eqref{eq_alpa} with $\alpha = -1.87 \pm 0.23$. \textbf{(b)} The number of infected individuals $I$ as a function of the city population, $m$. The scaling follows Eq. \eqref{eq_bet} with $\beta = 1.18 \pm 0.20$. \textbf{(c)} The scaling of $I$ with the distance-population ratio $r/m$ follows Eq. \eqref{eq:gamma_new} with $\gamma = -0.84 \pm 0.09$. The exponents within the errorbars follow Eq. \eqref{eq:exponents relation}. \textbf{(d)} The scaling of infected individuals with the population migration from Hubei. Almost linear scaling is observed  with $\phi = 0.96 \pm 0.25$ strongly relating the disease propagation to human mobility. The scaling of population migration with the \textbf{(e)} distance, \textbf{(f)} population and \textbf{(g)} distance-population ratio follow Eq. \eqref{eq:P_m} with $\tilde{\alpha} = -1.37 \pm 0.11$, $\tilde{\beta} = 1.11 \pm 0.09$ and $\tilde{\gamma} = -0.74 \pm 0.13$ respectively. The large value of $\alpha$ compared to $\tilde{\alpha}$ possibly indicate the quarantines efficiency.}
	\label{fig:spatial}	
	\end{center}
\end{figure*}
In the absence of both medicine and vaccination, strategies of effective distributing of them are not considered yet and the options to stop the propagation of the disease are currently to quarantines the infected individuals \cite{COVID19-li2020early} and social distancing \cite{gu2020mathematical} in order to cut the infection channels. Statistical estimations of the incubating (latency) period of the virus which includes no illness symptoms vary between different populations and found to be of about 4-6 days \cite{leung2020estimating} while a long incubation period of 19 days has also been observed \cite{bai2020presumed}. The 14 days quarantine period, which has been adopted by many countries, is a result of the high limit of the 95\% confidence levels \cite{linton2020incubation} of this incubating period. Under this quarantine strategy, the virus spreading could be alleviated. \par
The first wave of COVID-19 in China was controlled much faster compared to other locations in the world. Although China was the country with most infection cases up to the middle of March 2020 (Fig. \ref{fig:global_cases}a), it was able to stop the spreading while in other countries the disease kept propagating close to exponentially as can be seen in Fig. \ref{fig:global_cases}b. In fact, in most cities in China the spreading stopped approximately after 20 days as shown in Fig. \ref{fig:global_cases}c. In contrast, in other locations it got controlled very slowly as shown, e.g., for Italy in Fig. \ref{fig:global_cases}d. This suggests that one can learn from the disease decay in China and apply appropriate measures in other locations in the world.  While a general estimation of the disease evolution has been recently conducted \cite{maier2020effective,li2020note}, a comprehensive analysis of its spatio-temporal propagation which is important for epidemic forecast and modelling \cite{petropoulos2020forecasting,vespignani2020modelling,anastassopoulou2020data}, is still missing. In this manuscript, we study some aspects of the spatio-temporal propagation of the first wave of the COVID-19 virus and discuss the differences between China and other countries. The data source is available at \url{https://github.com/canghailan/Wuhan-2019-nCoV}.  \par
We study the spatial dynamics of the COVID-19 originated from Hubei and find scaling (power) laws for the number of infected individuals in different provinces as a function of the province population, the distance from Hubei and their relation to the population migration from Hubei. The human mobility \cite{gonzalez2008understanding,song2010modelling,gatto2020spread} which has been suggested to follow a L\'{e}vy-flight pattern \cite{brockmann2006scalinghumantravel,baronchelli2013levy,zaburdaev2015levy} is significantly important for modelling the spatial propagation of the disease. A reasonable explanation for the correlation between the population migration and the disease spread can be the strict quarantines applied in most cities in China after the shutdown of Wuhan traffic \cite{chinazzi2020effect,chen2020covid,lai2020effect}. These  quarantines were effective and probably prevented infected individuals to further spread the disease to other cities. Hence, the number of infected individuals is highly correlated to the population migration from Hubei before the shutdown, Fig. \ref{fig:spatial}. \par
As a result of the heterogeneous structure of cities in a country \cite{gross2020two}, quarantine has been usually applied strictly between cities and the mobility within cities was less restricted. Thus, more insight can be obtained in analyzing the epidemic within cities while studying the disease decay on a country scale might lead to uncertain conclusion regarding the disease situation since the disease may propagate in one city and decay in another. Our results suggest that the temporal propagation of the disease in most cities in China is similar. In fact, many cities in China experienced a two-stages process of the disease. At early times (of order of few days), the disease was undetectable due to the incubating period while spreading within the city. Another possibility is that it takes a few days for the city to become prepared for the disease. At later times, the infected individuals have been quarantined and the disease started to decay approximately exponentially in many cities. The quarantines have been effective to extinct the disease inside a city and reach a stable state with infection rates close to zero. Since quarantines were applied almost at the same time in most cities in China, we find that the decay stage of the disease starts almost at the same time for most of the cities no matter if they are large central cities, small cities or even Hubei province cities. For this reason most of the cities experience a similar 10-30 days characteristic time of the disease drastic reduction. In addition, most cities in China show similar exponential growth of the recovery rate which can reflect the similarity of health system efficiency. This in marked contrast to Italy which did not show an exponential growth probably due to overloaded health system by the many unexpected patients or because of a less obedient population to the health systems regulations. \par
\section{Spatial Scaling}
One of the most important properties of epidemics spreading is its spatial propagation, a characteristic which mainly depends on the epidemic mechanism, human mobility and control strategy. While the relation of human mobility to the epidemic spread has been shown \cite{kraemer2020effect,tian2020investigation}, a comprehensive picture of the epidemic spread in terms of distance, population size and human mobility and their scaling relations is still missing. 
We assume that the number of infected individuals in different provinces in China can be generally described as
\begin{equation}
I = f(r,m)
\label{general_I}
\end{equation}
where $r$ is the distance of the province from Wuhan and $m$ is the  population of the province. Since $r$ and $m$ are independent of each others, one can assume and study the scaling relation of each of them independently,
\begin{equation}
I \sim r^{\alpha}
\label{eq_alpa}
\end{equation}
and
\begin{equation}
I \sim m^{\beta} 
\label{eq_bet}.
\end{equation}
Since $r$ and $m$ are independent, Eq. \ref{general_I} can assume the scaling form (in analogous with population mobility \cite{simini2012universal}),
\begin{equation}
I \sim r^\nu / m^\mu
\label{I_spe}
\end{equation}
from which the relation $\nu/\alpha - \mu / \beta = 1$ should be satisfied. Using weighted least squares regression for the scaling we find in Fig. \ref{fig:spatial}a that $\alpha \simeq -1.87 \pm 0.23$ in agreement with a recent study \cite{biswas2020space} and $\beta \simeq 1.18 \pm 0.20$ as shown in Fig. \ref{fig:spatial}b. The minimization of the error of the exponents relation yields that $\nu = \mu \simeq -0.84 \pm 0.09$. Thus, Eq. \ref{I_spe} takes the form
\begin{equation}
I \sim (r/m)^\gamma
\label{eq:gamma_new}
\end{equation}
with $\gamma = \nu = \mu \simeq -0.84 \pm 0.09$ as shown in Fig. \ref{fig:spatial}c. Thus, $r/m$ can be regarded as a suitable distance-population parameter. The relation between these 3 exponents thus,
\begin{equation}
1/\alpha - 1 / \beta = 1/ \gamma .
\label{eq:exponents relation}
\end{equation}
\par
\begin{figure*}
	\begin{center}
	\includegraphics[scale=0.67]{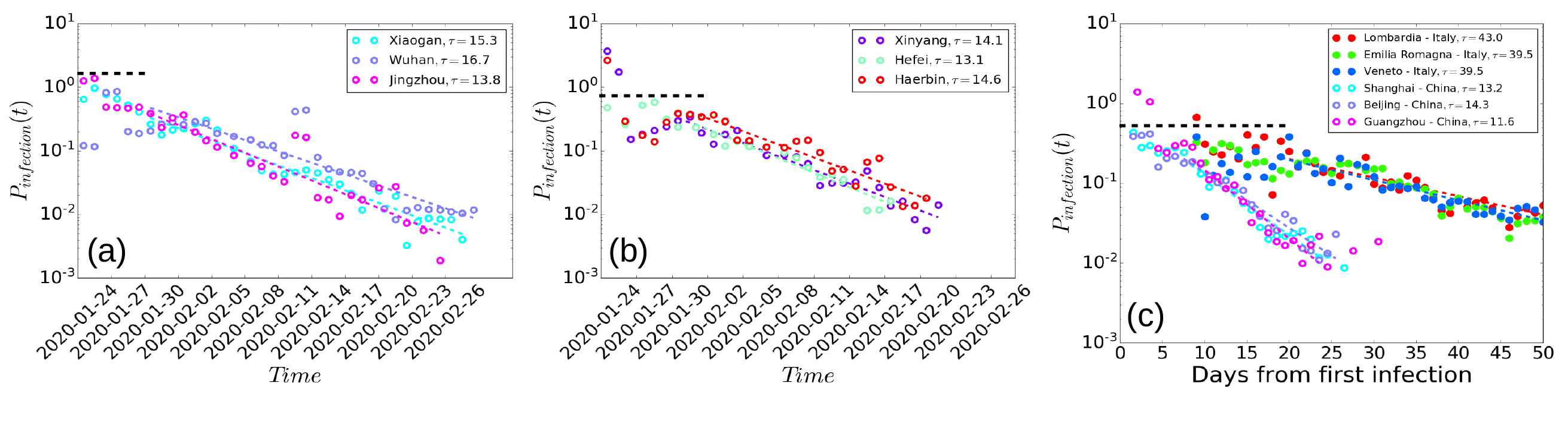}
	\caption{\textbf{Two-stages infection rate}. The infection rate $P_{infection}(t)$ for different types of cities: \textbf{(a)} cities in Hubei province, \textbf{(b)} small cities in China and \textbf{(c)} large central cities in China and provinces in Italy. At early times an approximate constant infection rate, $P_0 \sim 0.74 \pm 0.55$ is usually observed (below the black horizontal dashed line). After a few days, an exponential decay is observed in China representing the efficiency of the quarantines. The colored dashed lines are the best fit for the exponential decay, Eq. \ref{P_t_exp_decay}. The characteristic decay time $\tau$ in Eq. \ref{P_t_exp_decay} represents the time it takes to control the disease in a city. Interestingly, while the infection rate in cities (provinces) in both China and Italy is decaying, the characteristic decay parameter $\tau$ in Italy is several times longer than China due to different quarantine efficiency.}
	\label{fig:infection_rate}	
	\end{center}
\end{figure*}
In order to better understand the basic mechanism of the disease propagation, we examine the relation of the number of infected individuals in different provinces in China with the population migration, $P_m$, from Hubei. Our results shown in Fig. \ref{fig:spatial}d suggest an almost linear scaling relation
\begin{equation}
I \sim P^{\phi}_m
\label{eq:phi}
\end{equation}
with $\phi = 0.96 \pm 0.25$. This relation can be understood since strict quarantines were applied in most cities in China after the shutdown of Wuhan traffic. The quarantines were efficient to prevent infected individuals to spread the disease to other cities leading to a close to linear relation between the number of infected individuals and the population migration from Hubei. This supports the relation between the disease propagation and human mobility and indicates that earlier quarantine of Hubei could attenuate the world-wide spreading.  \par
To further study the relation between population migration and the disease propagation we measured the population migration number, $P_m$, as a function of the distance, population and the distance-population parameter. We assume the following scaling relations \cite{simini2012universal}, 
\begin{equation}
\begin{gathered}
P_m \sim r^{\tilde{\alpha}} \\
P_m \sim m^{\tilde{\beta}} \\
P_m \sim (r/m)^{\tilde{\gamma}} \\
\label{eq:P_m}
\end{gathered}
\end{equation}
with $\tilde{\alpha} = -1.37 \pm 0.11$, $\tilde{\beta} = 1.11 \pm 0.09$ and $\tilde{\gamma} = -0.74 \pm 0.13$ as shown in Figs. \ref{fig:spatial}e,f,g respectively. These exponents for the population migration represent the analogy of the exponents $\alpha$, $\beta$ and $\gamma$ of the number of infected individuals and follow a similar relation as Eq. \ref{eq:exponents relation} within the errorbars. Interestingly, $\tilde{\alpha}$ is lower than $\alpha$ and it can possibly be understood by quarantines efficiency which reduces the spatial spread of infected individuals compared to the population migration.  A summary of the exponents of \ref{eq_alpa}-\ref{eq:P_m} can be found in Table \ref{table:1}.  \par
\begin{table}[h!]
	\begin{center}
	\begin{tabular}{ | m{5em} | m{8em}| } 
		\hline
		\multicolumn{2}{|c|}{Spatial} \\
		\hline
		$\alpha$ & -1.87 $\pm$ 0.23\\
		$\tilde{\alpha}$ & -1.37 $\pm$ 0.11\\  
		\hline
		$\beta$ & 1.18 $\pm$ 0.20\\
		$\tilde{\beta}$ & 1.11 $\pm$ 0.09\\  
		\hline
		$\gamma$ & -0.84 $\pm$ 0.09\\ 
		$\tilde{\gamma}$ & -0.74 $\pm$ 0.13\\
		\hline
		$\phi$ & 0.96 $\pm$ 0.25\\ 
		\hline
		\multicolumn{2}{|c|}{Temporal} \\
		\hline
		$P_0$ & 0.74 $\pm$ 0.55\\ 
		\hline
		$\tau$ & 19.4 $\pm$ 8.3\\ 
		\hline
		$k$ & 0.027 $\pm$ 0.006\\
		\hline
	\end{tabular}
	\caption{\textbf{Spatio-temporal scaling exponents. Spatial -}  $\alpha$ is the exponent of the spatial distribution of infected individuals, Eq. \ref{eq_alpa}. $\beta$ is the scaling exponent of the population, $m$, Eq. \ref{eq_bet}, and $\gamma$ is the scaling exponent for the scaling function of the distance-population ratio $r/m$, Eq. \ref{eq:gamma_new}. The exponent $\phi$ relates the number of infected individuals to the measured number of population migration with almost linear scaling, Eq. \ref{eq:phi}. $\tilde{\alpha}$, $\tilde{\beta}$ and $\tilde{\gamma}$ are the exponents characterizing the scaling of the population migration with $r$, $m$ and $r/m$ respectively, Eq. \ref{eq:P_m}. \textbf{Temporal -} $P_0$ is the approximately constant infection rate at early times while the disease is spreading. $\tau$ is the characteristic time of the disease decay at later times, assuming exponential decay, Eq. \ref{P_t_exp_decay}. $k$ is the growth parameter of the recovery rate, Eq. \ref{recovery_rate}.}
	\label{table:1}
	\end{center}
\end{table}
\section{Temporal behaviour}
The absence of vaccination makes the control of the disease very difficult and the main action possible is to quarantine infected individuals and those that were in contact with them in order to prevent further spreading. This approach is effective but limited since an infected individual can spread the disease before showing illness symptoms. Nonetheless, an efficient quarantine strategy can succeed in controlling the disease and a method to quantify its efficiency is needed. In addition, while a recent study showed that quarantine was efficient resulting in a subexponential growth of the confirmed cases in different cities in China \cite{maier2020effective}, it did not show how it affects the infection rate which is important for modeling. Moreover, a temporal analysis of the death and recovery rates is required as they may be affected by the health system efficiency which might be overloaded. Here we will analyze the temporal evolution of the infection, death, and recovery rates under the quarantine restrictions in different cities in China  and compare it to different provinces in Italy. \par
Since quarantines have been usually applied within and between cities, studying the disease decay on a country scale might lead to uncertain conclusion regarding the disease situation since the disease may propagate in one city and decay in another. Thus, to further study the effect of quarantines, we measured the infection rate in different cities in China and different provinces in Italy. The infection rate, $P_{infection}(t)$, is measured for each city (province) using the total number of infected individuals in the city in a given day, $I(t)$, from the first day that infected individuals have been detected in the city. The infection rate at a given day is defined as the fraction of newly infected individuals emerging from the total number of infected individuals a day earlier
\begin{equation}
P_{infection}(t) = \frac{I(t) - I(t-1)}{I(t-1)} .
\label{P_t1}
\end{equation}
We examined three different types of cities in China. $a)$ cities in Hubei province, $b)$ small cities and $c)$ large central cities as shown in Fig. \ref{fig:infection_rate}a,b,c respectively. In all three cases, an approximately constant infection rate is observed in early times. However, after a few days, a decay in the infection rate is observed. Determining if the decay is exponential or power-law can not be certain due to the few data points in the samples. Assuming exponential decay \cite{pastor2001epidemic} gives a plausible consistent picture. In this case, Eq. \ref{P_t1} takes the form
\begin{equation}
P_{infection}(t) = \begin{cases}
P_0 & t_0 < t < t_x\\
P_0e^{-(t - t_x)/ \tau} & t_x < t ,
\end{cases}
\label{P_t_exp_decay}
\end{equation}
where $P_0$ is the constant infection rate without constrains, $t_0$ represents the time that the first infected individual was detected in the city, $t_x$ is the time when the quarantine starts and $\tau$ is the characteristic time for the disease drastic reduction.
The approximately constant infection rate in early times represents the real infection rate of the disease before quarantines were applied to control the disease while the exponential decay in later times represents the efficiency of quarantines in reducing the infected rate. Small value of the parameter $\tau$ indicate more efficient restrictions. The constant value $t_x$ is very similar in different cities in China due to the similar emergency response of other provinces with respect to the epidemic outbreak. The exponential decay with low values of $\tau$ indicates that quarantines are efficient to tame the disease in most cities in China and indeed, in the last days of February, the infection rate has been almost zero with rarely new cases as seen in Fig. \ref{fig:infection_rate}. In marked contrast, the decay in Italy has been slower with much larger values of $\tau$ indicating less efficient quarantine strategy as seen in Fig. \ref{fig:infection_rate}c. While the infection rate might be biased by different number of tests and reporting policies \cite{piguillem2020optimal}, an efficient quarantine should still result with a decay of the infection rate.\par
\begin{figure}
	\includegraphics[scale=0.83]{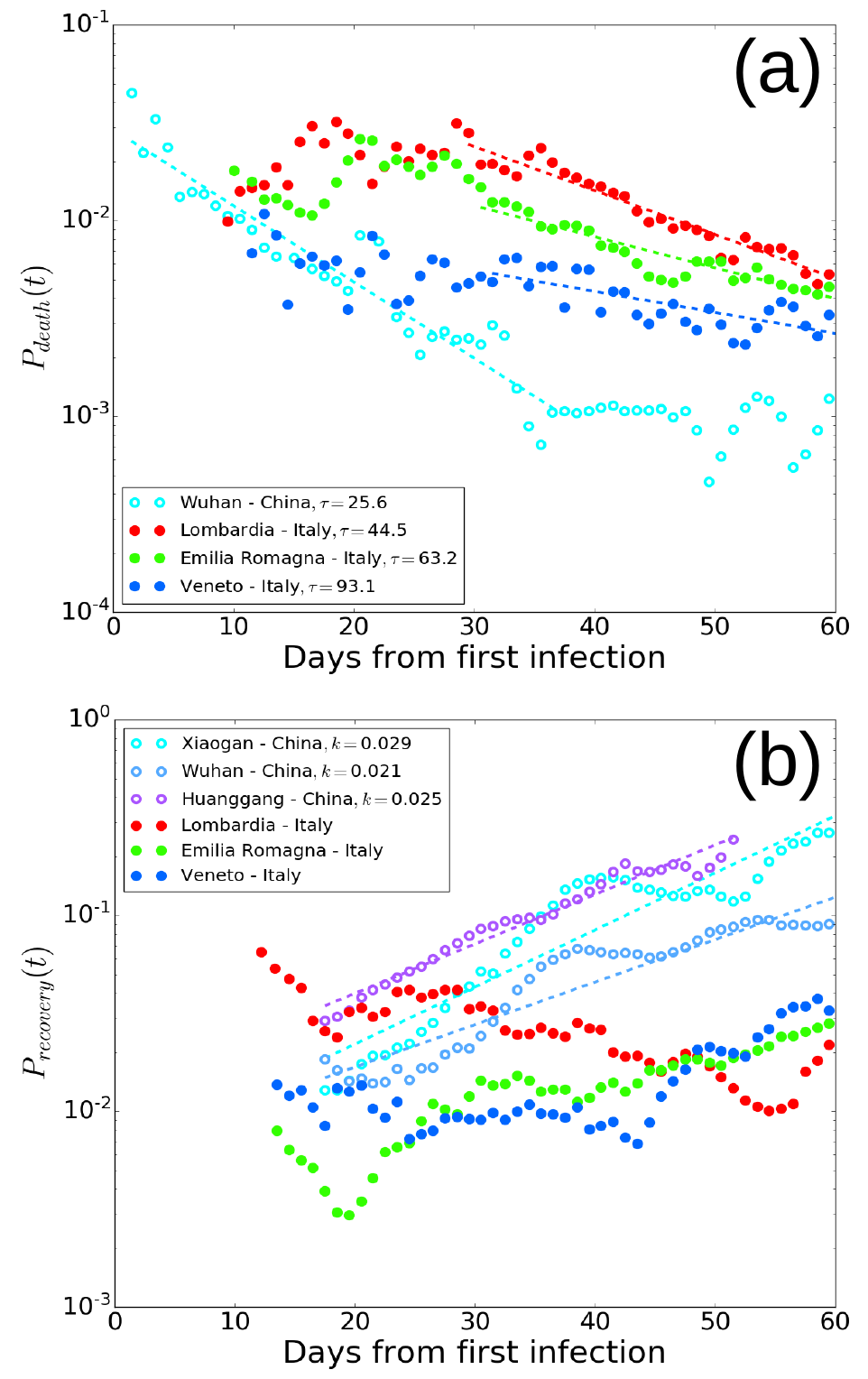}
	\caption{\textbf{Death and recovery rates. (a)} The death rate in Wuhan shows an approximate exponential decay similar to the infection rate (other cities show poor statistics). The characteristic decay time $\tau$ is about twice time longer for the death rate than the infection rate. Provinces in Italy with most cases show much slower decay suggesting an overloaded health system. \textbf{(b)} The recovery rate in China shows an approximate exponential growth as the disease is getting controlled, Eq. \ref{recovery_rate}. The growth parameter $k$ characterizes the health system efficiency. Provinces in Italy with most recovery cases do not show growth and remain approximately constant suggesting an overloaded health system as suggested by the death rate.}
	\label{fig:recovery_death}	
\end{figure}
While the infection rate characterize the quarantine efficiency, the death and recovery rates can characterize the health system efficiency which may be overloaded by the unexpected amount of patients. The death and recovery rates are defined as the fraction of the newly dead and recovered individuals at each day and the number of infected individual a days earlier
\begin{equation}
P_{death}(t) = \frac{D(t) - D(t-1)}{I(t-1)} 
\label{death_rate}
\end{equation}
and
\begin{equation}
P_{recovery}(t) = \frac{R(t) - R(t-1)}{I(t-1)} 
\end{equation}
where $D(t)$ is the number of dead people at time $t$ and $R(t)$ is the number of recovered at time $t$.
The death rate in Wuhan shows an approximate exponential decay similar to the infection rate with characteristic time close to twice longer. In marked contrast, the provinces with most cases in Italy show much slower decay, as seen in Fig. \ref{fig:recovery_death}a, which indicates an overloaded health system. \par
The recovery rate in China increases exponentially as the disease gets controlled with less new cases and can be well approximated  as 
\begin{equation}
P_{recovery}(t) \sim e^{kt},
\label{recovery_rate}
\end{equation}
where $k$ is the growth parameter which might indicate the health system efficiency. In Italy the recovery rate is approximately constant, as shown in Fig. \ref{fig:recovery_death}b, which suggests, similarly to the death rate, an overloaded health system as shown in Fig. \ref{fig:recovery_death}a. \par
\begin{figure}
	\begin{center}
	\includegraphics[scale=0.9]{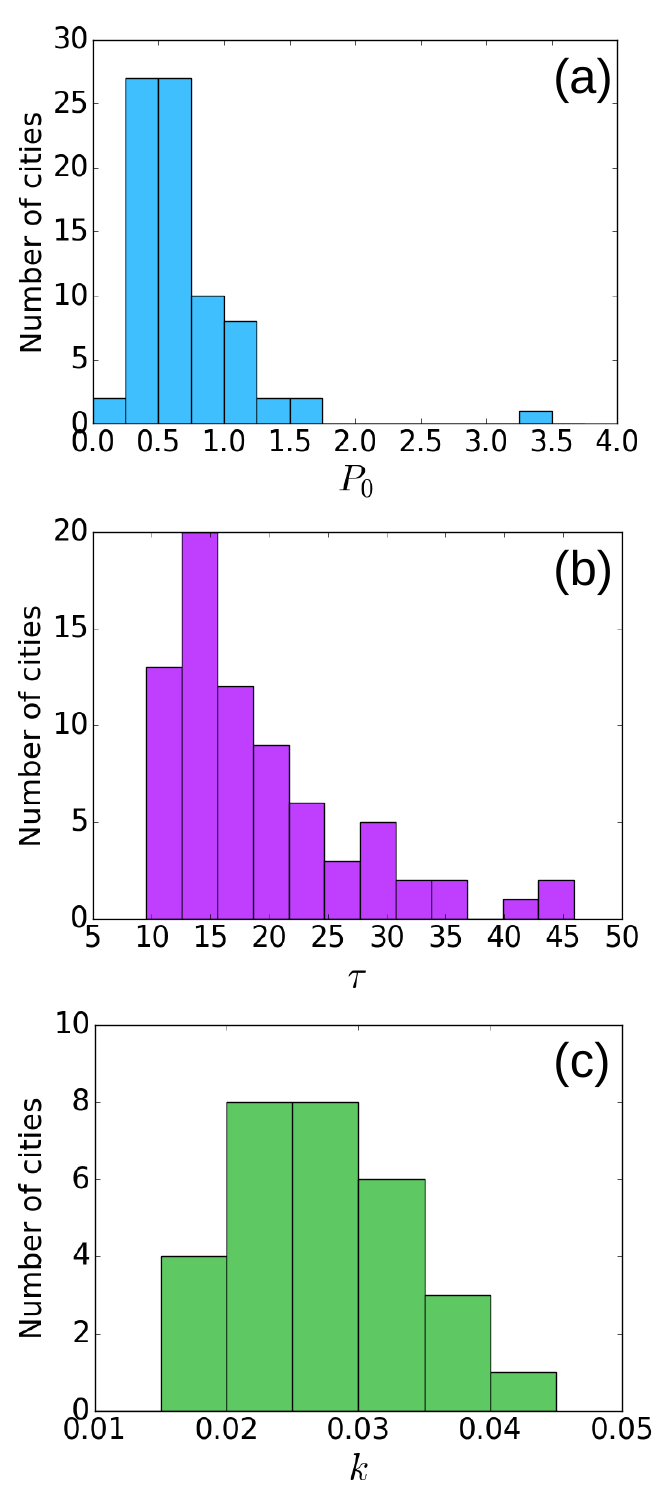}
	\caption{\textbf{Statistical properties of the temporal behaviour of cities in China} \textbf{(a)}  The constant infection rate, $P_0$ is mostly in the range 0-2 while in a few cities it is much larger. The average value is $0.74$ with standard deviation $0.55$. \textbf{(b)} The distribution of the exponent $\tau$ characterizing the extinction of the disease which found to be in the range 10-50 days. The average value is $19.4$ with standard deviation $8.3$. \textbf{(c)} The distribution of $k$ characterizing the recovery rate and reflects the health system efficiency is found to be in the range between 0.015-0.045. The average value is $0.027$ with standard deviation $0.006$. The parameters are very similar in most cities in China consistent with the strict quarantine applied in them simultaneously.}
	\label{fig:P_t_hist}
	\end{center}
\end{figure}

The temporal parameters characterizing the infection, death and recovery rates are found to be similar for most cities in China with $P_0$ in the range 0.5-2 with an average of $0.74$ and a standard deviation of $0.55$ as shown in Fig. \ref{fig:P_t_hist}a.
%Considering that an individual is detected after 5-14 days of being infected and within $l$ days it infect approximation $P_0$ people, it will infect 1-3 people. 
The value of $\tau$ for most of the cities is 10-30 days while for a few cities the characteristic time can be longer as seen in Fig. \ref{fig:P_t_hist}b. The value of $\tau$ may characterizes the efficiency of quarantines. The average value of $\tau$ is $19.4$ with standard deviation $8.3$. The values of $k$ are in range 0.015 - 0.045 and may characterize the health system efficiency. The average value of $k$ is $0.027$ with standard deviation $0.006$. Summary of the temporal parameters can be found in Table \ref{table:1}. The similar values of the temporal parameters in most cities in China is consistent with the strict quarantine applied in them simultaneously.\par
\section{Discussion and summary}
While many countries have been struggling to overcome the first wave of the  COVID-19, China controlled it relatively fast. Controlling the disease is not easy, the spread of the disease is highly related to population migration which assumed to follows a L\'{e}vy flight behavior, a characteristic of human mobility with long jumps which spread the disease rapidly. The incubating period together with the L\'{e}vy flight long jumps makes the disease very hard to control since by the time that infected individual is being detectable, it can already perform a long-distance trip and further spread the disease. This spatial dynamics is very important and should be taken into account in epidemic modeling. Despite these difficulties, an efficient quarantine strategy could control the disease leading to an exponential decay of the infection rate. This decay is characterized by the characteristic decay time $\tau$ and allows a quantitative comparison between quarantine strategies performed in different places. In fact, the lifetime of the disease in a city is characterized by two stages, uncontrolled infection in early times (for few days) and decaying stage at later times once quarantines start to effect. These two stages can explain the temporal dynamics of the disease situation in China in the first wave and explain the situation in other locations in the world where similar strategies have been only partly adopted. Moreover, this two-stage process is very important for modeling since one should take into account how the infection rate changes in time due to quarantine strategies. \par
In addition, while the temporal dynamics of the infection rate is related to the quarantine efficiency, the temporal dynamics of the death and recovery rates is more related to the health system efficiency which may be overloaded or inefficient. Even though different countries are characterized by different age population distribution which suffers from different mortality rates \cite{dowd2020demographic}, a sharp rise of the death rate (or very slow decay) may indicate an overloaded health system. The same conclusion applies for the case of a decay of the recovery rate.\par
Our work highlights the importance of efficient quarantine strategies and the strong relation between population migration and the disease spreading. Moreover, our results suggest that early action may attenuate the disease propagation and prevent overload of the health system which have not been ready for the large amount of unexpected new patients. \par
\acknowledgments
We thank Ivan Bonamassa for very useful discussions related to this project. We thank the Israel Science Foundation (Grant no. 189/19) and the joint China-Israel Science Foundation (Grant no. 3132/19), ONR, the BIU Center for Research in Applied Cryptography and Cyber Security, and DTRA Grant no. HDTRA-1-19-1-0016 for financial support.

\bibliographystyle{unsrt}
\bibliography{mybib}
\end{document}